%% file: paper.tex
\documentclass[10pt]{article}

\usepackage{amsmath}
\usepackage{url,hyperref}

\input JackMacros

\input{flatex}

\input{macros}

\usepackage{booktabs} 

\newcommand{\NoShow}[1]{}

\begin{document}
\title{Deriving Correct High-Performance Algorithms\\[0.2in]
\large
FLAME Working Note \#86
}

\author{Devangi N. Parikh\\ Margaret E. Myers\\ Robert A. van de Geijn\\
\normalsize
The University of Texas at Austin \\
\normalsize
Austin, Texas 78712 \\
\texttt{\{dnp,myers,rvdg\}@cs.utexas.edu}
}

\maketitle

\begin{abstract}
Dijkstra observed that verifying correctness of a program is difficult and conjectured that derivation of a program hand-in-hand with its proof of correctness was the answer.
We illustrate this goal-oriented approach by applying it to 
the domain of dense linear algebra libraries for distributed memory parallel computers.  
We show that
algorithms that underlie the implementation of most functionality for this domain can be systematically derived to be correct.  The benefit is that an entire family of algorithms for an operation is discovered so that the best algorithm for a given architecture
can be chosen.   This approach is very practical: Ideas inspired by it have been used to rewrite the dense linear algebra software stack starting below the Basic Linear Algebra Subprograms (BLAS) and reaching up through the Elemental distributed memory library, and every level in between.  
The paper demonstrates how formal methods and rigorous mathematical techniques for correctness impact 
HPC. 
\end{abstract}

\input{body}

\newpage

\bibliographystyle{plain}
\bibliography{biblio} 

\end{document}

%% file: JackMacros.tex
\usepackage{tikz}
\usepackage{epsfig}
\usepackage{amssymb}
\usepackage{program}
\usepackage{ifthen}
\usepackage{color}
\usepackage{boxedminipage}
\usepackage{fancyvrb}
\usepackage[dotinlabels]{titletoc}
\usepackage{colortbl}
\usepackage{array}
\usepackage{soul}
\usepackage{graphicx}
\usepackage{graphics}
\usepackage{moreverb}
\usepackage{rotating}
\usepackage{enumerate}
\usepackage{pgfplots}
\pgfplotsset{compat=newest}

\pgfplotscreateplotcyclelist{journal}{%
blue,every mark/.append style={fill=blue!80!black},mark=square*\\%
red,densely dashed,every mark/.append style={solid,fill=red!80!black},mark=o\\%
brown!80!black,every mark/.append style={fill=brown!80!black},mark=diamond\\%
green,every mark/.append style={fill=green!80!black},mark=diamond*\\%
cyan,densely dashed,every mark/.append style={solid,fill=cyan!80!black},mark=*\\%
magenta,densely dashed,every mark/.append style={solid,fill=magenta!80!black},mark=square*\\%
green,densely dashed,every mark/.append style={solid,fill=red!80!black},mark=o\\%
}

\newcommand{\stack}[2]{
\mathord{
  \raise 1.2ex\hbox to 0pt 
          {$\scriptstyle#2$\hss}
  #1}}

\newcommand{\colv}[1]{
\mathord{
  \raise 1.8ex\hbox to 0pt 
          {$\scriptstyle \downarrow$\hss}
  #1}}



%% file: flatex.tex
\newcolumntype{I}{!{\vrule width 1.5pt}}
\newlength\savedwidth
\newcommand\whline{\noalign{\global\savedwidth\arrayrulewidth
                            \global\arrayrulewidth 1.5pt}%
           \hline
           \noalign{\global\arrayrulewidth\savedwidth}}

\newboolean{IsWide}
\setboolean{IsWide}{true}

\newcommand{\FlaTwoByTwo}[4]{
\left( 
\begin{array}{c I c}
#1 & #2 \\ \whline
#3 & #4 
\end{array} 
\right)
}

\newcommand{\FlaThreeByThreeTL}[9]{
\left( 
\begin{array}{c | c I c}
#1 & #2 & #3 \\ \hline
#4 & #5 & #6 \\ \whline
#7 & #8 & #9
\end{array} 
\right) 
}

\newcommand{\FlaThreeByThreeBR}[9]{
\left( 
\begin{array}{c I c | c}
#1 & #2 & #3 \\ \whline
#4 & #5 & #6 \\ \hline
#7 & #8 & #9
\end{array} 
\right)
}

\newcommand{\FlaPartition}[2]{
\ifthenelse{\boolean{IsWide}}{{\bf partition } \hspace{-1em} #1 \hspace{-1em} #2}
{{\bf partition } \+ \\ #1 \+ \\ #2 \- \-}
}

\newcommand{\FlaRepartition}[2]{
\ifthenelse{\boolean{IsWide}}{{\bf repartition } \hspace{-1em} #1 \hspace{-1em} #2}
{{\bf repartition } \+ \\ #1 \+ \\ #2 \- \-}
}

\newcommand{\FlaContinue}[1]{
\ifthenelse{\boolean{IsWide}}{{\bf continue with } #1
}
{{\bf continue with } \+ \\ #1 \-
}
}

\newcommand{\FlaStartComputeShorter}{
\setlength{\unitlength}{0.5in}
\begin{picture}(3,0.01)
\put(0,0){\line(1,0){4}}
\put(0,0.01){\line(1,0){4}}
\end{picture} 
}

\newcommand{\FlaEndComputeShorter}{
\setlength{\unitlength}{0.5in}
\begin{picture}(3,0.01)
\put(0,0){\line(1,0){4}} 
\put(0,0.01){\line(1,0){4}} 
\end{picture} 
}

\newcommand{\operation}{ [ D, E, F, \ldots ] \becomes {\rm op}( A, B, C, D, \ldots ) }
\newcommand{\routinename}{ [ D, E, F, \ldots ] \becomes {\rm op}( A, B, C, D, \ldots ) }
\newcommand{\routinecost}{ X }
\newcommand{\precondition}{ Q }
\newcommand{\postcondition}{ R }
\newcommand{\invariant}{ P }
\newcommand{\costinv}{ \  }

\newcommand{\guard}{ R }
\newcommand{\partitionings}{
\begin{minipage}{2in}
$ S_I $
\end{minipage}
}
\newcommand{\partitionsizes}{ \hspace{ 1.25in} }

\newcommand{\blocksize}{1}

\newcommand{\repartitionings}{
\begin{minipage}[t]{2in}
\ \\
\ \\
\ \\
\end{minipage}
}

\newcommand{\repartitionsizes}{ \hspace{ 1.25in} }
\newcommand{\moveboundaries}{
\begin{minipage}[t]{2in}
\ \\
\ \\
\ \\
\end{minipage}
}
\newcommand{\beforeupdate}{
$ \QBefore $
}
\newcommand{\afterupdate}{
$ \QAfter $
}
\newcommand{\update}{%
\begin{minipage}[t]{4in}
$ S_U $
\end{minipage}
}

\newcommand{\resetsteps}{
\renewcommand{\blocksize}{1}
\renewcommand{\operation}{ [ D, E, F, \ldots ] \becomes {\rm op}( A, B, C, D, \ldots ) }
\renewcommand{\routinename}{ [ D, E, F, \ldots ] \becomes {\rm op}( A, B, C, D, \ldots ) }
\renewcommand{\routinecost}{ 0 }
\renewcommand{\precondition}{ \PPre }
\renewcommand{\postcondition}{ \PPost }
\renewcommand{\invariant}{ \PInv }
\renewcommand{\costinv}{ \  }
\renewcommand{\guard}{ G }
\renewcommand{\partitionings}{ %
\begin{minipage}[t]{2in}
\ \\
\end{minipage}
}
\renewcommand{\partitionsizes}{ \hspace{ 2.25in} }
\renewcommand{\repartitionings}{%
\begin{minipage}[t]{2in}
\ \\
\end{minipage}
}
\renewcommand{\repartitionsizes}{ \hspace{ 1.25in} }
\renewcommand{\moveboundaries}{%
\begin{minipage}[t]{2in}
\ \\
\end{minipage}
}
\renewcommand{\beforeupdate}{
\QBefore
}
\renewcommand{\afterupdate}{
\QAfter
}
\renewcommand{\update}{
$ S_U $
}
}

\newcommand{\WSprecondition}{
$ \left\{ \precondition \right\} $
}

\newcommand{\WSpostcondition}{
$ \left\{ \postcondition \right\} $
}

\newcommand{\WSinvariant}{
$ \left\{ \invariant \right\} $
}

\newcommand{\WStopofloop}{
$ \left\{ \left( \invariant \right) \wedge \left( \guard \right) \right\} $
}

\newcommand{\WSafterloop}{
$ \left\{ \left( \invariant \right) \wedge \neg \left( \guard \right) \right\} $
}

\newcommand{\WSguard}{
$ \guard $
}

\newcommand{\WSpartition}{
\begin{minipage}[t]{2in}
\begin{tabbing}
ind \= ind \= \kill
{\bf Partition} 
\partitionings{\bf where } \partitionsizes 
\end{tabbing}
\end{minipage}
}

\newcommand{\WSpartitionNarrow}{
\begin{minipage}[t]{2.2in}
\begin{tabbing}
ind \= ind \= \kill
{\bf Partition} 
\partitionings \+ \\
{\bf where } \hspace*{-2ex} \partitionsizes\vspace{-0.2in}
\end{tabbing}
\end{minipage}
}

\newcommand{\WSrepartition}{
\begin{minipage}[t]{5.0in}
\ifthenelse{ \equal{\blocksize}{1} }{}
{%
\ifthenelse{ \equal{\blocksize}{2} }{~}
{\bf Determine blocksize} $ \blocksize $ \\
}
{\bf Repartition}
\begin{tabbing}
in \= in \= \+ \kill
\repartitionings
\\
{\bf where } \hspace*{-2ex} \repartitionsizes 
\end{tabbing}
\end{minipage}
}

\newcommand{\WSrepartitionNarrow}{
\begin{minipage}[t]{2.15in}
\ifthenelse{ \equal{\blocksize}{1} }{}
{%
\ifthenelse{ \equal{\blocksize}{blank} }{~}
{\bf Determine block size $ \blocksize $} \\
}
{\bf Repartition}
\begin{tabbing}
i \= i \= \+ \kill
\repartitionings \+ \\
{\bf where } \hspace*{-2ex} \repartitionsizes 
\end{tabbing}
\end{minipage}
}

\newcommand{\WSmoveboundary}{
\begin{minipage}[t]{6.0in}
{\bf Continue with} \\
\moveboundaries 
\end{minipage}
}

\newcommand{\WSmoveboundaryNarrow}{
\begin{minipage}[t]{2.15in}
{\bf Continue with}
\begin{tabbing}
i \= \+ \kill
\moveboundaries 
\end{tabbing}
\end{minipage}
}

\newcommand{\WSbeforeupdate}{
$ \left\{ \beforeupdate \right\} $
}

\newcommand{\WSafterupdate}{
$ \left\{ \afterupdate \right\} $
}

\newcommand{\WSupdate}{
\update
}

\newcommand{\worksheet}{
\begin{center}
\begin{tabular}{| c I l |} \hline
\footnotesize {\bf Step} & 
{\bf Annotated Algorithm:} $\operation$ 
\\ \whline 
\rowcolor[gray]{0.93}
1a & 
\WSprecondition \\ \hline
4 & 
\WSpartition \\ \hline
\rowcolor[gray]{0.93}
2 & 
\WSinvariant \\ \hline
3 & 
{\bf while} \WSguard { \bf do} \\ \hline
\rowcolor[gray]{0.93}
2,3 &
\ \hspace{0.15in} \WStopofloop \\ \hline
5a & 
\ \hspace{0.15in} 
\WSrepartition
\\ \hline
\rowcolor[gray]{0.93}
6 &
\ \hspace{0.15in} \WSbeforeupdate \\ \hline
8 & \ \hspace{0.15in} \WSupdate \\ \hline
5b &
\ \hspace{0.15in} 
\WSmoveboundary
\\ \hline
\rowcolor[gray]{0.93}
7 &
\ \hspace{0.15in} \WSafterupdate \\ \hline 
\rowcolor[gray]{0.93}
2 &
\ \hspace{0.15in} \WSinvariant \\ \hline
& {\bf endwhile} \\ \hline
\rowcolor[gray]{0.93}
2,3 &
\WSafterloop \\ \hline
\rowcolor[gray]{0.93}
1b &
\WSpostcondition \\ \hline
\end{tabular}
\end{center}
}

\newcommand{\FlaAlgorithmNarrow}{
\begin{center}
\begin{tabular}{|@{\hspace{3pt}}l@{\hspace{2pt}} |} \hline
{\bf Algorithm:} $\routinename$ 
\\ \whline
{\WSpartitionNarrow} \\
{\bf while} \WSguard { \bf do} \\
\ \hspace{0.0in} \WSrepartitionNarrow \\
{\hspace{0.0in} \FlaStartComputeShorter} \\
{\hspace{0.0in} \WSupdate} \\
{\hspace{0.0in} \FlaEndComputeShorter} \\
{\ \hspace{0.0in} \WSmoveboundaryNarrow} \\
{{\bf endwhile}} \\ \hline
\end{tabular}
\end{center}
}

\newcommand{ \PPre }{ P_{\it pre} }
\newcommand{ \PPost }{ P_{\it post} }
\newcommand{ \PInv }{ P_{\it inv} }

\newcommand{ \becomes }{:=}
\newcommand{ \QBefore }{ P_{\it before} }
\newcommand{ \QAfter }{ P_{\it after} }

%% file: macros.tex
\newcounter{mycounter}



%% file: body.tex
\section{Introduction}

\input{intro_final}
\label{sec:intro}

\section{Deriving from specification}
\label{sec:derivation}

\input{alg}

\NoShow{
\section{Discussion}
\label{sec:Discussion}

\input discussion

}

\section{Questions and answers}

\input questions







\section{Conclusion}
\label{sec:conclusion}

\input conclusion

\vspace{0.1in}
\noindent
{\bf Acknowledgments.}
This research was supported in part by NSF Award ACI-1550493.  We would like to thank our many collaborators who have contributed to this research over the years.

\noindent
{\em Any opinions, findings, and conclusions or recommendations expressed in this material are those of the author(s) and do not necessarily reflect the views of the National Science Foundation.}

%% file: intro_final.tex
A typical approach to the porting of functionality to a new architecture is to retarget an existing implementation. Along the way, a software developer may use a profiler to identify the compute-intensive part of the code, which then becomes the first step in a long optimization process.
Sometimes (often?), after a lot of effort, it is discovered that an entirely new algorithm should be used instead.   

As an example, let us consider an important step when computing the solution of the generalized Hermitian eigenvalue problem.  
This problem can be formulated mathematically as
$Ax=\lambda Bx$, where $A$ is Hermitian
and $B$ is Hermitian and positive-definite.  
The positive definiteness of
$ B $ can be exploited by computing its  Cholesky factorization ($ B = L L^H $ where $ L $ is lower triangular)
and transforming the problem into a 
standard Hermitian eigenvalue problem:
$ L^{-1} A L^{-H} z = \lambda z $
where $ z = L^H x $.

While high-performance implementation of Cholesky factorization is a well-studied topic~\cite{Bientinesi:2008:FAR:1377603.1377606}, 
the computation 
$ A = L^{-1} A L^{-H} $, also called  {\em two-sided triangular solve with multiple right-hand sides} (two-sided {\sc trsm}) since it can also be reformulated as solving 
$
 L C L^{H} = A $,
where $ C $ overwrites $ A $,
is an important operation that is a bit more off the beaten track.

The distributed memory parallel implementation of this operation appears to have started with the direct translation of the LAPACK~\cite{LAPACK3} routine {\sc [sdcz]sygst} into the ScaLAPACK routine
{\sc p[sdcz]sygst}%
\footnote{This is an example that we encountered many years ago.  We consider it a convenient example to make our point.  What we describe about the history of this  operation should not be taken to be criticism of current implementations.  The performance graphs we use were taken from a technical report written many years ago.  The point of this paper is to answer the question of how to make the discovery of algorithms systematic.  The performance graphs used are merely there for illustration.  The focus of the reader should be on how deriving correct algorithms helps discover high-performance solutions.}.  
Since much of the computation in {\sc [sdcz]sygst} involves a triangular solve with multiple right-hand sides operation {\sc [sdcz]trsm}, which is known not to parallelize well, the resulting performance was extremely poor, as illustrated in Figure~\ref{fig:perf:sygst_all_top} (the performance curve is among those at the very bottom).  A better implementation, which uses a totally different algorithmic variant, proposed by Sears et al.~\cite{Sears98}, yields much better performance (given by the curve {\tt pdsyngst}) and is also available in ScaLAPACK.  The associated paper describes the new algorithm, but does not describe how to systematically find it:  It is due to the special skill of a few experts that  a better algorithm was found.  The technique is special enough that the better algorithm was instantiated for the case where (due to symmetry) only the lower triangular part of matrix $ A $ is updated, but not for the case where the upper triangular part is targeted instead.

\input perf_fig_sygst_rl.tex

In Figure~\ref{fig:perf:sygst_all_top}, we also show performance curves for five more algorithms.  They are implemented using the Elemental library for distributed memory parallel dense matrix computions.  The point is that there are many algorithms, among which is an algorithm that best translates to distributed memory architectures.  The details of the optimization of the chosen algorithm are secondary to the choice of that algorithm.  Our approach is captured by a quote attributed to Dijkstra: ``Always design your program as a member of a whole family of programs, including those that are likely to succeed it.''  

\NoShow{
What if we change how we develop code?  What if we start with a careful specification of the operation to be computed rather than a legacy implementation?  What if we can systematically derive a family of algorithms from specification, from which to choose the most appropriate for a given target architecture?  What if we can then systematically explore {\em all} optimizations of the chosen algorithm, guided by analytical models?  What if we can teach this to a relative novice?  What if all of this can be made mechanical (automated)?  In other words: What if we elevate software development from an art to a science?

This survey paper narrates what has been accomplished by us and others over the past two decades towards the goal of systematically deriving algorithms.
}

%% file: perf_fig_sygst_rl.tex
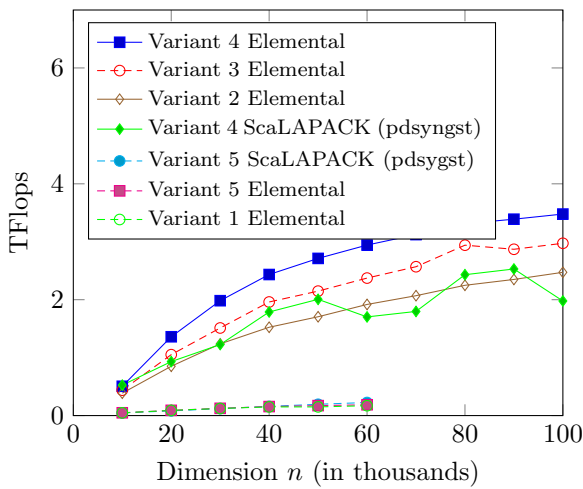
\begin{figure}[tb!]
\begin{center}
\begin{tikzpicture}
  \begin{axis}[width=0.455\textwidth,
               xlabel=Dimension $n$ (in thousands),
               ylabel={TFlops},
               xmin=0,xmax=100,ymin=0,ymax=7,
               legend style={legend pos=north west},
               clip=false,cycle list name=journal]
  \addplot coordinates {
      (10,0.50568) 
      (20,1.36006) 
      (30,1.98273) 
      (40,2.4346)  
      (50,2.71269) 
      (60,2.94299) 
      (70,3.12362) 
      (80,3.31129) 
      (90,3.38956) 
      (100,3.47825) 
  };
  \addplot coordinates {
      (10,0.43811) 
      (20,1.05155) 
      (30,1.51192) 
      (40,1.96052) 
      (50,2.14809) 
      (60,2.37313) 
      (70,2.56909) 
      (80,2.94094) 
      (90,2.87032) 
      (100,2.97376) 
  };
  \addplot coordinates {
      (10,0.38645) 
      (20,0.8511)  
      (30,1.23789) 
      (40,1.52426) 
      (50,1.70552) 
      (60,1.91802) 
      (70,2.06905) 
      (80,2.24849) 
      (90,2.34966) 
      (100,2.4727) 
  };
  \addplot coordinates {
      (10,0.52554) 
      (20,0.93173) 
      (30,1.22594) 
      (40,1.78805) 
      (50,2.00619) 
      (60,1.70291) 
      (70,1.79758) 
      (80,2.43142) 
      (90,2.53007) 
      (100,1.97784) 
  };
  \addplot coordinates {
      (10,0.04645) 
      (20,0.08089) 
      (30,0.12063) 
      (40,0.15873) 
      (50,0.19517) 
      (60,0.22579) 
  };
  \addplot coordinates {
      (10,.0481)
      (20,.09037)
      (30,.12527)
      (40,.15562)
      (50,.17178)
      (60,.18629)
  };
  \addplot coordinates {
      (10,.04695)
      (20,.08956)
      (30,.1229)
      (40,.14974)
      (50,.14974)
      (60,.17023)
  };
  \legend{
\begin{minipage}{1.75in}\footnotesize Variant 4 Elemental\end{minipage},
\begin{minipage}{1.75in}\footnotesize Variant 3 Elemental\end{minipage},
\begin{minipage}{1.75in}\footnotesize Variant 2 Elemental\end{minipage},
\begin{minipage}{1.75in}\footnotesize Variant 4 ScaLAPACK (pdsyngst)\end{minipage},
\begin{minipage}{1.75in}\footnotesize Variant 5 ScaLAPACK (pdsygst)\end{minipage},
\begin{minipage}{1.75in}\footnotesize Variant 5 Elemental\end{minipage},
\begin{minipage}{1.75in}\footnotesize Variant 1 Elemental\end{minipage},
}
  \end{axis}
\end{tikzpicture}
\end{center}
\caption{Performance of the various implementations 
on  2048 cores of Blue Gene/P.
The top of the graph represents the theoretical peak of this architecture.
(The three curves for Variants~1 and~5, which cast substantial computation
in terms of a parallel {\sc trsm}, essentially coincide near
the bottom of the graph.)
The legend lists the implementations from fastest to slowest for the largest
problem size.
Results courtesy of Dr. Jack Poulson.
For details on how these experiments were conducted, consult~\cite{FLAWN56}.
}
\label{fig:perf:sygst_all_top}
\end{figure}

%% file: alg.tex
``The only effective way to raise the confidence level of a program significantly is to give a convincing proof of its correctness. But one should not first make the program and then prove its correctness, because then the requirement of providing the proof would only increase the poor programmer's burden. On the contrary: the programmer should let correctness proof and program grow hand in hand.''
- Edsger W. Dijkstra (1972)~\cite{EWD:EWD340}

In this section, we show how  goal-oriented programming, advocated by Dijkstra and others, can be made practical for HPC.
The methodology was first proposed in a dissertation by Gunnels~\cite{Gunnels:PhD} and a related paper~\cite{FLAME}. It was refined into the  ``worksheet'' that is explained in this section, in~\cite{Recipe}.

We focus on the two-sided TRSM as the driving example: Solve
\[
 L C L^{H} = A ,
\quad
  \mbox{overwriting $ A $ with the solution $ C $.}
 \]
Since $ A $ is symmetric, we will only use and update its lower triangular part.

We give the information required so that those
familiar with the FLAME methodology~\cite{Gunnels:2001:FFL,Recipe,Bientinesi2013,TSoPMC,LAFF-On-edX,LAFF-On} can understand how the algorithms 
were derived while others will at least appreciate the high level ideas.

\subsection{Derivation of a family of algorithms}

We reformulate the computation $ A \becomes L^{-1} A L^{-H} $
as the constraint 
$ A = C \wedge L C L^H = \hat{A} $ where $ \wedge $ denotes the logical AND
operator.
This constraint expresses
that $ A $ is to be overwritten by matrix $ C $, where $ C $ satisfies
the given constraint in which $ \hat{A} $ represents the input matrix
$ A $.  This constraint is known as the {\em postcondition} in the FLAME
methodology.
Details, such as the fact that $ A $ is symmetric and that only the lower triangular part is to be updated, are kept implicit in our discussion.
Deriving algorithms now becomes an eight step process, captured in Figure~\ref{fig:WS}.  This ``worksheet'' is filled in the order indicated in the left column marked ``Step.''

\vspace{0.05in}
\noindent
{\bf Step 1: Precondition and postcondition.}
We start by entering the {\em precondition} (the state of the variables at the start) and {\em postcondition} (the state of the variables upon completion) in the worksheet.

\vspace{0.05in}
\noindent
{\bf Step 2: Derivation of the loop-invariants.}
Key to proving a loop correct is an assertion known as the {\em loop-invariant}.  By showing that {\em if} it holds before an iteration, then it holds again after the iteration, mathematical induction tells us it holds for all iterations.  
The question is how to systematically derive the loop-invariant so it can used to derive the loop.
A recursive definition of the operation, which we call the {\em Partitioned Matrix Expression} (PME), provides the answer.

We form the PME
by partitioning the matrices so that
\[
\setlength{\arraycolsep}{2pt}
A \rightarrow
\FlaTwoByTwo 
 { A_{TL} }{ \star }
 { A_{BL} }{ A_{BR} },
C \rightarrow
\FlaTwoByTwo 
 { C_{TL} }{ \star }
 { C_{BL} }{ C_{BR} },
\mbox{ and }
L \rightarrow
\FlaTwoByTwo 
 { L_{TL} }{ 0 }
 { L_{BL} }{ L_{BR} },
\]
where $ A_{TL} $, $ C_{TL} $, and $ L_{TL} $ are square submatrices and
$ \star $ denotes the parts of the Hermitian matrices that are neither stored nor updated.
Partitioning captures that algorithms inherently ``march'' through matrices in a way that identifies regions that have or have not been (partially) updated and/or which which computation has been performed.  

Substituting these partitioned matrices into the postcondition yields
\begin{eqnarray*}
	\setlength{\arraycolsep}{2pt}
	\footnotesize
\FlaTwoByTwo 
 { A_{TL} }{ \star }
 { A_{BL} }{ A_{BR} }
= 
\FlaTwoByTwo 
 { C_{TL} }{ \star }
 { C_{BL} }{ C_{BR} }
\wedge 
	\setlength{\arraycolsep}{2pt}
	\footnotesize
& \hspace{-0.1in}
	\setlength{\arraycolsep}{2pt}
	\footnotesize
\begin{array}[t]{c}
\underbrace{
\FlaTwoByTwo 
 { L_{TL} }{ 0 }
 { L_{BL} }{ L_{BR} }
\FlaTwoByTwo 
 { C_{TL} }{ \star }
 { C_{BL} }{ C_{BR} }
\FlaTwoByTwo 
 { L_{TL} }{ 0 }
 { L_{BL} }{ L_{BR} }^H
= 
\FlaTwoByTwo 
 { \hat A_{TL} }{ \star }
 { \hat A_{BL} }{ \hat A_{BR} }
} \\
\footnotesize
\FlaTwoByTwo
  { L_{TL} C_{TL} L_{TL}^H = \hat A_{TL} }{ \star }
  { L_{BR} C_{BL}  = \hat A_{BL} L_{TL}^{-H} - L_{BL} C_{TL} }
  { 
\begin{array}[t]{r c l}
L_{BR} C_{BR} L_{BR}^H = \hat A_{BR} - L_{BL} C_{TL} L_{BL}^H  \\
 ~~~-  L_{BL} C_{BL}^H L_{BR}^H  
  - L_{BR} C_{BL} L_{BL}^H 
\end{array}
 }.
\end{array}
\end{eqnarray*}
This expresses all conditions that must be satisfied upon completion of the
computation, in terms of the submatrices.
The bottom-right quadrant can be further manipulated into
\begin{eqnarray*}
\lefteqn{L_{BR} C_{BR} L_{BR}^H = 
\hat A_{BR} - L_{BL} C_{TL} L_{BL}^H -  L_{BL} C_{BL}^H L_{BR}^H 
 - L_{BR} C_{BL} L_{BL}^H } \\
&\hspace{-0.05in}=
\footnotesize
\hat A_{BR} 
\!-\!
L_{BL}
\begin{array}[t]{c}
\underbrace{
\left( 
\frac{1}{2} C_{TL} L_{BL}^H  
+ C_{BL}^H L_{BR}^H 
\right) }
\\ 
W_{BL}^H
\end{array} 
\!\!-\!\!
\begin{array}[t]{c}
\underbrace{
\left( \frac{1}{2} L_{BL} C_{TL}
+ L_{BR} C_{BL} \right)  }
\\
W_{BL}
\end{array}
L_{BL}^H
\end{eqnarray*}
using a standard trick to cast three rank-$k$ updates into 
a single symmetric rank-$2k$ update.
Thus, the PME can be rewritten as
\begin{eqnarray*}
\setlength{\arraycolsep}{3pt}
\FlaTwoByTwo 
 { A_{TL} }{ \star }
 { A_{BL} }{ A_{BR} }
= 
\FlaTwoByTwo 
 { C_{TL} }{ \star }
 { C_{BL} }{ C_{BR} }
&\wedge &
Y_{BL} = L_{BL} C_{TL} 
\wedge
W_{BL} = L_{BR} C_{BL} - \frac{1}{2} Y_{BL}
\\
& \wedge &
\FlaTwoByTwo
  { L_{TL} C_{TL} L_{TL}^H = \hat A_{TL} }{ \star }
  { L_{BR} C_{BL} \!=\! \hat A_{BL} L_{TL}^{-H} \!-\! Y_{BL} }
  { 
  	\begin{array}[t]{l}
  		L_{BR} C_{BR} L_{BR}^H \\
  		~~~~ = \hat A_{BR} \!-\! ( L_{BL} W_{BL}^H \!+\! W_{BL} L_{BL}^H ) 
  		\end{array}}.
\end{eqnarray*}

\input invariants

\input GenEigTransformWS
\input GenEigTransformNew

We are now ready to identify  loop-invariants for algorithms.
In the case of this operation, there are many such loop invariants.
However, careful consideration for maintaining symmetry in the
intermediate update and avoiding unnecessary computation leaves the five tabulated
in Figure~\ref{fig:invariants}.  What would have made Dijkstra happy with us is that we determine loop invariants {\em a priori}.

To now derive a specific algorithm, one picks one of the loop invariants for the remainder of the steps.  In our discussion, we pick the simplest, Invariant 1.

\vspace{0.05in}
\noindent
{\bf Steps~3 and~4: Loop guard and initialization.}
The next step is to determine under what condition, the {\em loop guard}, execution of the loop is not yet finished.  Determining this condition means reasoning as follows:  After the loop completes, the loop guard is {\em false} and the loop invariant still holds.  This must imply that we have completed the correct result.  It is not hard to see that when $ A_{TL} $, $ C_{TL}$, etc., become all of $ A $, $ C $, etc. (respectively), the loop invariant implies the postcondition.  

Similarly, before we start the loop, the loop invariant must hold.  Thus, we need to partition the matrices in such a way that the precondition implies the loop invariant.  Step 4 in Figure~\ref{fig:WS} has this property.

\vspace{0.05in}
\noindent
{\bf Step~5: Progress.}
Next we observe that progress must be made through the matrices.  In other words, the top-left quadrants $ A_{TL} $, $ C_{TL} $, etc., must expand.  We denote this  by repartitioning, exposing submatrices (in order to derive high-performance {\em blocked} algorithms) that are taken from some of the quadrants at the top of the loop and added to other quadrants at the bottom of the loop.

\vspace{0.05in}
\noindent
{\bf Step 6: State after repartitioning.}
The algorithm must have the property that it maintains Invariant~1.
We know what the state of the various quadrants of $ A $ is at the top of the loop.  The {\bf repartitioning} step is really an indexing step: it exposes submatrices without updating their contents.  Some of these submatrices will need to be updated in order to maintain the loop invariant.  Step~6 answers the question ``What is the state of the matrix $ A $ at the top of the loop in terms of the exposed blocks?''  The answer can be  systematically determined by reasoning through how the exposed submatrices are related to the quadrants that appear in the loop invariant.

\vspace{0.05in}
\noindent
{\bf Step~7: State after moving the computation forward.}
Again, the algorithm must have the property that it maintains Invariant~1.
We know what the state of the various quadrants of $ A $ is at the bottom of the loop.  The {\bf continue with} step is again an indexing step: it consolidates submatrices into quadrants without updating their contents.  Step~7 answers the question ``What is the state of the matrix $ A $ at the bottom of the loop in terms of the exposed blocks?''  Again, the answer can be systematically determined by reasoning through how the exposed submatrices are related to the quadrants that appear in the loop invariant.

\vspace{0.05in}
\noindent
{\bf Step~8: Updating the exposed submatrices.}
Finally, we recognize that the contents of the submatrices must be updated from the state indicated in Step~6 to the state indicated in Step~7.  

\vspace{0.05in}
\noindent
{\bf The algorithms.}
The eight steps now yield the algorithmic {\em variant}, blocked Variant~1, that corresponds to Invariant~1.  All the assertions in the grey boxes in the worksheet provide the proof of correctness, which has been created hand-in-hand with the derivation of the algorithm.  After eliminating those grey boxes, we are left with the algorithm, as shown in Figure~\ref{fig:algs}.  In that figure, we also give Variants~2-5, which result from applying the same steps to Invariants~2-5.

\subsection{Not all algorithms are created equal}
All algorithms in Figure~\ref{fig:algs} incur a cost of about $ n^3 $ floating point operations (flops) when $ n $ is the matrix size.
A quick way to realize where the algorithms 
spend most of their time is to consider the partitionings
\[
\FlaThreeByThreeBR
  { \cellcolor[gray]{0.8} A_{00} }{ \star }{ \star }
  { A_{10} }{ A_{11} }{ \star }
  { \cellcolor[gray]{0.8} A_{20} }{ A_{21} }{ \cellcolor[gray]{0.8} A_{22} }
,
\FlaThreeByThreeBR
  { \cellcolor[gray]{0.8} L_{00} }{ 0 }{ 0 }
  { L_{10} }{ L_{11} }{ 0 }
  { \cellcolor[gray]{0.8} L_{20} }{ L_{21} }{ \cellcolor[gray]{0.8}
  L_{22} },
\mbox{ and }
\FlaThreeByThreeBR
  { \cellcolor[gray]{0.8} C_{00} }{ \star }{ \star }
  { C_{10} }{ C_{11} }{ \star }
  { \cellcolor[gray]{0.8} C_{20} }{ C_{21} }{ \cellcolor[gray]{0.8}
  C_{22} }, 
\] 
and to note that operations that involve at least one
operand that is highlighted contribute to an $ O(n^3) $ (highest order) 
cost term while the others contribute to at most an $ O( b n^2 ) $
term.  Thus, first and foremost, it is important that the
highlighted operations in Figure~\ref{fig:algs} 
attain high
performance.

On sequential and shared memory parallel architectures, all of the operations highlighted in Figure~\ref{fig:algs} can be implemented with level-3 BLAS (matrix-matrix) calls, which in principle can attain high performance~\cite{BLAS3,Goto1,Goto2}.
In practice, there are some differences.
These differences become very pronounced on parallel architectures, as illustrated in Figure~\ref{fig:perf:sygst_all_top}.

As was already pointed out in a paper by Sears et al.~\cite{Sears98},
and is explored in more detail in~\cite{FLAWN56},
it is the parallel triangular solves with $ b $ right-hand
sides ({\sc trsm}), 
$ A_{10} \becomes A_{10} L_{00}^{-H} $ in Variant~1 and 
$ A_{21} \becomes L_{22}^{-1} A_{21} $ in Variant~5, that inherently
do {\em not} parallelize well yet account for about 1/3 of the flops
for Variants~1 and~5.
The reason is that inherent dependencies exist within the
{\sc trsm} operation,
the details of which go beyond the scope of this paper.  All of the other
highlighted operations can, in principle, asymptotically attain
near-peak performance when correctly parallelized on an architecture
with reasonable communication~\cite{SUMMA,pblas3,IPPS:98,PLAPACK}.  
Thus, Variants~1 and~5 cast a substantial
fraction of computation in terms of an operation that does not
parallelize well, in contrast to Variants~2, 3, and~4.
Variant~3 has the disadvantage that intermediate result $ Y_{BL} $ must be 
stored.  An expert knows that (general of Hermitian) rank-k updates (the case of matrix-matrix multiplication where the ``k'' dimension is small) inherently parallelizes well.  It is for this reason that Variant 4 comes out on top.
Variant 2 might be a good choice when implementing an out-of-core
algorithm, since the highlighted computations for it 
require the bulk of data ($ A_{00} $ and $ A_{20} $) to be read
but not written.  In other words, different circumstances call for different algorithmic variants.

%% file: invariants.tex
\begin{figure}[tbp]
\renewcommand*\arraystretch{1.4}
\begin{center}
\begin{tabular}{| l |} \hline
\underline{Loop Invariant 1}  
\\
\footnotesize
{\setlength{\arraycolsep}{2pt}
$
 \FlaTwoByTwo 
 { A_{TL} }{ \star }
 { A_{BL} }{ A_{BR} }
=
\FlaTwoByTwo 
 { C_{TL} }{ \star }
 { \hat{A}_{BL} 
}
 { \hat A_{BR} 
 }
 \wedge L_{TL} C_{TL} L_{TL}^{-1} = \widehat A_{TL}
$
}
\\ \hline\underline{Loop Invariant 2}  
\\
\footnotesize
{\setlength{\arraycolsep}{2pt}
$
 \FlaTwoByTwo 
 { A_{TL} }{ \star }
 { A_{BL} }{ A_{BR} }
=
\FlaTwoByTwo 
 { C_{TL} }{ \star }
 { \hat{A}_{BL} L_{TL}^{-H} }
 { \hat A_{BR} 
 }
 \wedge L_{TL} C_{TL} L_{TL}^{-1} = \widehat A_{TL}
$
}
\\ \hline
\underline{Loop Invariant 3} \\
\footnotesize
{\setlength{\arraycolsep}{2pt}
$
\begin{array}{@{}l}
 \FlaTwoByTwo 
 { A_{TL} }{ \star }
 { A_{BL} }{ A_{BR} }
=
\FlaTwoByTwo 
 { C_{TL} }{ \star }
 { \hat A_{BL} L_{TL}^{-H} }
 { \hat A_{BR} 
 } 
\wedge L_{TL} C_{TL} L_{TL}^{-1} = \widehat A_{TL}\\
~~~~~~~~~~~~~~~ \wedge 
 \FlaTwoByTwo 
 { Y_{TL} }{ }
 { Y_{BL} }{ Y_{BR} }
= 
 \FlaTwoByTwo 
 {  }{ }
 { L_{BL} C_{TL} }{ ~~~~~~ }
 \end{array}
$
}
\\ \hline
\underline{Loop Invariant 4} 
\\ 
{
\footnotesize
{\setlength{\arraycolsep}{2pt}
$
\begin{array}{@{}l}
 \FlaTwoByTwo 
 { A_{TL} }{ \star }
 { A_{BL} }{ A_{BR} }
=
\FlaTwoByTwo 
 { C_{TL} }{ \star }
 { \hat A_{BL} L_{TL}^{-H} - L_{BL} C_{TL} }
 { \hat A_{BR} 
        - ( L_{BL} W_{BL}^H + W_{BL} L_{BL}^H ) 
 } \\
 ~~~~~~~~~~~
 \wedge \cdots
 \end{array}
$
}
}
\\ \hline
\underline{Loop Invariant 5} 
\\ 
{
\footnotesize
{\setlength{\arraycolsep}{2pt}
$
 \FlaTwoByTwo 
 { A_{TL} }{ \star }
 { A_{BL} }{ A_{BR} }
=
\FlaTwoByTwo 
 { C_{TL} }{ \star }
 { C_{BL} }
 { \hat A_{BR} 
        - ( L_{BL} W_{BL}^H + W_{BL} L_{BL}^H ) 
 }
 \wedge \cdots
$
}
}
\\ \hline
\end{tabular}
\end{center}
\caption{Five loop invariants for computing $ A \becomes L^{-1} A L^{-H} $.}
\label{fig:invariants}
\end{figure}

%% file: GenEigTransformWS.tex
\resetsteps      


\renewcommand{\operation}{ A = C \wedge L C L^H = \widehat A }

\renewcommand{\routinename}{\operation}


\renewcommand{\precondition}{
	A = \widehat{A}
}


\renewcommand{\postcondition}{ 
A = C \wedge L C L^H = \widehat A
}


\renewcommand{\invariant}{
	 \FlaTwoByTwo 
	 { A_{TL} }{ \star }
	 { A_{BL} }{ A_{BR} }
	 =
	 \FlaTwoByTwo 
	 { C_{TL} }{ \star }
	 { \hat{A}_{BL} 
	 }
	 { \hat A_{BR} 
	 }
	 \wedge L_{TL} C_{TL} L_{TL}^{-1} = \widehat A_{TL}
}


\renewcommand{\guard}{
	m( A_{TL} ) < m( A )
}


\renewcommand{\partitionings}{
	$
	A \rightarrow
	\FlaTwoByTwo{A_{TL}}{A_{TR}}
	{A_{BL}}{A_{BR}}
	$
	,
	$
	L \rightarrow
	\FlaTwoByTwo{L_{TL}}{L_{TR}}
	{L_{BL}}{L_{BR}}
	$
	,
	$
	Y \rightarrow
	\FlaTwoByTwo{Y_{TL}}{Y_{TR}}
	{Y_{BL}}{Y_{BR}}
	$ \\
}

\renewcommand{\partitionsizes}{
	$ A_{TL} $ is $ 0 \times 0 $,
	$ L_{TL} $ is $ 0 \times 0 $,
	$ Y_{TL} $ is $ 0 \times 0 $
}


\renewcommand{\blocksize}{b}

\renewcommand{\repartitionings}{
	$  \FlaTwoByTwo{A_{TL}}{A_{TR}}
	{A_{BL}}{A_{BR}}
	\rightarrow
	\FlaThreeByThreeBR{A_{00}}{A_{01}}{A_{02}}
	{A_{10}}{A_{11}}{A_{12}}
	{A_{20}}{A_{21}}{A_{22}}
	$,
	$  \FlaTwoByTwo{L_{TL}}{L_{TR}}
	{L_{BL}}{L_{BR}}
	\rightarrow
	\FlaThreeByThreeBR{L_{00}}{L_{01}}{L_{02}}
	{L_{10}}{L_{11}}{L_{12}}
	{L_{20}}{L_{21}}{L_{22}}
	$, \\
	$  \FlaTwoByTwo{Y_{TL}}{Y_{TR}}
	{Y_{BL}}{Y_{BR}}
	\rightarrow
	\FlaThreeByThreeBR{Y_{00}}{Y_{01}}{Y_{02}}
	{Y_{10}}{Y_{11}}{Y_{12}}
	{Y_{20}}{Y_{21}}{Y_{22}}
	$}

\renewcommand{\repartitionsizes}{
	$ A_{11} $ is $ b \times b $,
	$ L_{11} $ is $ b \times b $,
	$ Y_{11} $ is $ b \times b $}


\renewcommand{\moveboundaries}{
	$  \FlaTwoByTwo{A_{TL}}{A_{TR}}
	{A_{BL}}{A_{BR}}
	\leftarrow
	\FlaThreeByThreeTL{A_{00}}{A_{01}}{A_{02}}
	{A_{10}}{A_{11}}{A_{12}}
	{A_{20}}{A_{21}}{A_{22}}
	$,
	$  \FlaTwoByTwo{L_{TL}}{L_{TR}}
	{L_{BL}}{L_{BR}}
	\leftarrow
	\FlaThreeByThreeTL{L_{00}}{L_{01}}{L_{02}}
	{L_{10}}{L_{11}}{L_{12}}
	{L_{20}}{L_{21}}{L_{22}}
	$, \\
	$  \FlaTwoByTwo{Y_{TL}}{Y_{TR}}
	{Y_{BL}}{Y_{BR}}
	\leftarrow
	\FlaThreeByThreeTL{Y_{00}}{Y_{01}}{Y_{02}}
	{Y_{10}}{Y_{11}}{Y_{12}}
	{Y_{20}}{Y_{21}}{Y_{22}}
	$}


\renewcommand{\beforeupdate}{
	\FlaThreeByThreeBR{A_{00}}{A_{01}}{A_{02}}
	{A_{10}}{A_{11}}{A_{12}}
	{A_{20}}{A_{21}}{A_{22}}
	=
	\FlaThreeByThreeBR{C_{00}}{\widehat A_{01}}{\widehat A_{02}}
	{\widehat A_{10}}{\widehat A_{11}}{\widehat A_{12}}
	{\widehat A_{20}}{\widehat A_{21}}{\widehat A_{22}}
	\wedge
	L_{00} C_{00} L_{00}^H = \widehat A_{00}
}


\renewcommand{\afterupdate}{
	\begin{array}{l}
	\FlaThreeByThreeTL{A_{00}}{A_{01}}{A_{02}}
	{A_{10}}{A_{11}}{A_{12}}
	{A_{20}}{A_{21}}{A_{22}}
	=
	\FlaThreeByThreeTL{C_{00}}{\widehat A_{01}}{\widehat A_{02}}
	{C_{10}}{C_{11}}{\widehat A_{12}}
	{\widehat A_{20}}{\widehat A_{21}}{\widehat A_{22}} \\
	\hspace{0.5in} \wedge
	\begin{array}[t]{l} 
		\underbrace{
		\begin{array}{c | c}
		L_{00} C_{00} L_{00}^H = \widehat A_{00} & \star \\ \hline
		( L_{10} C_{00} + L_{11} C_{10}) L_{00}^H= \widehat A_{10} &
		L_{10} C_{00} L_{10}^H + 
		L_{11} C_{10} L_{10} + L_{10} C_{10}^H L_{11}^H
		+ L_{11} C_{11} L_{11}^H = \widehat A_{11}
	\end{array}
}\\ 
Y_{10} = L_{10} C_{00} \wedge
W_{10} = L_{11} C_{10} - \frac{1}{2} Y_{10} 
\wedge \\
	\begin{array}[t]{c | c}
	L_{00} C_{00} L_{00}^H = \widehat A_{00} & \star \\ \hline
 L_{11} C_{10} = \widehat A_{10} L_{00}^{-H} - Y_{10} &
	L_{11} C_{11} L_{11}^H = \widehat A_{11} - ( L_{10} W_{10}^H + W_{10} L_{10}^H)
\end{array}
\end{array}
\end{array}
}


\renewcommand{\update}{
	$
	\begin{array}[t]{l}
	\rowcolor[gray]{0.8}    
	Y_{10} \becomes L_{10} A_{00} ~~~ \mbox{({\sc hemm})} \\
	\rowcolor[gray]{0.8}    
	A_{10} \becomes A_{10} L_{00}^{-H} ~~~ \mbox{({\sc trsm})} \\
	\rowcolor[gray]{1.0} 
	A_{10} \becomes W_{10} = A_{10} - \frac{1}{2} Y_{10} \\
	A_{11} \becomes A_{11} - ( A_{10} L_{10}^H + L_{10} A_{10}^H) \\
	A_{11} \becomes L_{11}^{-1} A_{11} L_{11}^{-H}\\
	A_{10} \becomes A_{10} - \frac{1}{2} Y_{10}\\
	A_{10} \becomes L_{11}^{-1} A_{10}
	\end{array}
	$
}

\begin{figure*}[tbp]
\setlength{\arraycolsep}{3pt}
\footnotesize
\begin{center}
\worksheet
\end{center}

\caption{Derivation of blocked Variant 1.
}
\label{fig:WS}
\end{figure*}

%% file: GenEigTransformNew.tex
\resetsteps      


\renewcommand{\routinename}{ A := L^{-1} A L^{-H}}


\renewcommand{\guard}{
  m( A_{TL} ) < m( A )
}


\renewcommand{\partitionings}{
  $
  A \rightarrow
  \FlaTwoByTwo{A_{TL}}{A_{TR}}
              {A_{BL}}{A_{BR}}
  $
,
  $
  L \rightarrow
  \FlaTwoByTwo{L_{TL}}{L_{TR}}
              {L_{BL}}{L_{BR}}
  $
,
  $
  Y \rightarrow
  \FlaTwoByTwo{Y_{TL}}{Y_{TR}}
              {Y_{BL}}{Y_{BR}}
  $
}

\renewcommand{\partitionsizes}{
$ A_{TL} $,
$ L_{TL} $, and $ Y_{TL} $ are $ 0 \times 0 $.
}


\renewcommand{\blocksize}{b}

\renewcommand{\repartitionings}{
$
  \FlaTwoByTwo{A_{TL}}{ \star }
              {A_{BL}}{A_{BR}}
  \rightarrow
  \FlaThreeByThreeBR{A_{00}}{ \star }{ \star }
                    {A_{10}}{A_{11}}{ \star }
                    {A_{20}}{A_{21}}{A_{22}}
$,
$
  \FlaTwoByTwo{L_{TL}}{ 0 }
              {L_{BL}}{L_{BR}}
  \rightarrow
  \FlaThreeByThreeBR{L_{00}}{0}{0}
                    {L_{10}}{L_{11}}{0}
                    {L_{20}}{L_{21}}{L_{22}}
$, \\
$
  \FlaTwoByTwo{Y_{TL}}{ 0 }
              {Y_{BL}}{Y_{BR}}
  \rightarrow
  \FlaThreeByThreeBR{Y_{00}}{0}{0}
                    {Y_{10}}{Y_{11}}{0}
                    {Y_{20}}{Y_{21}}{Y_{22}}
$
}

\renewcommand{\repartitionsizes}{
  $ A_{11} $,
  $ L_{11} $, and $ Y_{11} $ are $ b \times b $
}


\renewcommand{\moveboundaries}{
$
  \FlaTwoByTwo{A_{TL}}{ \star }
              {A_{BL}}{A_{BR}}
  \leftarrow
  \FlaThreeByThreeTL{A_{00}}{ \star }{ \star }
                    {A_{10}}{A_{11}}{ \star }
                    {A_{20}}{A_{21}}{A_{22}}
$,
$
  \FlaTwoByTwo{L_{TL}}{0}
              {L_{BL}}{L_{BR}}
  \leftarrow
  \FlaThreeByThreeTL{L_{00}}{0}{0}
                    {L_{10}}{L_{11}}{0}
                    {L_{20}}{L_{21}}{L_{22}}
$, \\
$
  \FlaTwoByTwo{Y_{TL}}{0}
              {Y_{BL}}{Y_{BR}}
  \leftarrow
  \FlaThreeByThreeTL{Y_{00}}{0}{0}
                    {Y_{10}}{Y_{11}}{0}
                    {Y_{20}}{Y_{21}}{Y_{22}}
$}

\renewcommand{\update}{
	$
	\begin{array}{l}
\\
%
%
    A_{10} \becomes L_{11}^{-1} A_{10} \\
%
%
A_{20} \becomes A_{20} - L_{21} A_{10} ~~ \mbox{({\sc gemm})} \\
%
%
    A_{11} \becomes L_{11}^{-1} A_{11} L_{11}^{-H} \\
%
%
    Y_{21} \becomes L_{21} A_{11} \\
%
%
    A_{21} \becomes A_{21} L_{11}^{-H} \\
    A_{21} \becomes W_{21} = A_{21} - \frac{1}{2} Y_{21} \\
%
%
A_{22} \becomes A_{22} - ( L_{21} A_{21}^H + A_{21} L_{21}^H ) ~~ \mbox{({\sc her2k})} \\
    A_{21} \becomes A_{21} - \frac{1}{2} Y_{21}  \\
  \end{array}
  $
}

\renewcommand{\update}{
\begin{tabular}{ @{} l | @{\hspace{2pt}} l |  @{\hspace{2pt}} l } 
	\underline{Variant 1} 
	&
	\underline{Variant 2}
	& 
	\underline{Variant 3}
	\\
	$  \begin{array}[t]{l}
	\rowcolor[gray]{0.8}    
	Y_{10} \becomes L_{10} A_{00} ~~~ \mbox{({\sc hemm})} \\
	\rowcolor[gray]{0.8}    
	A_{10} \becomes A_{10} L_{00}^{-H} ~~~ \mbox{({\sc trsm})} \\
	\rowcolor[gray]{1.0} 
	A_{10} \becomes W_{10} = A_{10} - \frac{1}{2} Y_{10} \\
	A_{11} \becomes A_{11} \\
	- ( A_{10} L_{10}^H + L_{10} A_{10}^H) \\
	A_{11} \becomes L_{11}^{-1} A_{11} L_{11}^{-H}\\
	A_{10} \becomes A_{10} - \frac{1}{2} Y_{10}\\
	A_{10} \becomes L_{11}^{-1} A_{10}
	\end{array}
	$ 
	&
	$
	\begin{array}[t]{l}
	\rowcolor[gray]{0.8}    
	Y_{10} \becomes L_{10} A_{00} ~~~ \mbox{({\sc hemm})} \\
	\rowcolor[gray]{1.0} 
	A_{10} \becomes W_{10} = A_{10} - \frac{1}{2} Y_{10} \\
	A_{11} \becomes A_{11} \\
	- ( A_{10} L_{10}^H + L_{10} A_{10}^H )  \\
	A_{11} \becomes L_{11}^{-1}  A_{11} L_{11}^{-H} \\
	\rowcolor[gray]{0.8}    
	A_{21} \becomes A_{21} - A_{20} L_{10}^H ~~~ \mbox{({\sc gemm})} \\
	\rowcolor[gray]{1.0} 
	A_{21} \becomes A_{21} L_{11}^{-H} \\
	A_{10} \becomes A_{10} - \frac{1}{2}  Y_{10} \\
	A_{10} \becomes L_{11}^{-1} A_{10} \\
	\end{array}
	$
	&
	$
	\begin{array}[t]{l}
	A_{10} \becomes W_{10} = A_{10} - \frac{1}{2} Y_{10} \\
	A_{11} = A_{11} \\
	- ( A_{10} L_{10}^H + L_{10} A_{10}^H ) \\
	A_{11} = L_{11}^{-1} A_{11} L_{11}^{-H}\\
	\rowcolor[gray]{0.8}    
	A_{21} = A_{21} - A_{20} L_{10}^H ~~~ \mbox{({\sc gemm})} \\
	\rowcolor[gray]{1.0} 
	A_{21} = A_{21} L_{11}^{-H} \\
	A_{10} = A_{10} - \frac{1}{2} Y_{10} \\
	A_{10} = L_{11}^{-1} A_{10}\\
	\rowcolor[gray]{0.8}        
	Y_{20} = Y_{20} + L_{21} A_{10} ~~~ \mbox{({\sc gemm})}    \\
	\rowcolor[gray]{1.0} 
	Y_{21} = L_{21} A_{11}\\
	\rowcolor[gray]{0.8}    
	Y_{21} = Y_{21} + L_{20} A_{10}^H ~~~ \mbox{({\sc gemm})}
	\end{array}
	$
	\\
	\hline
	\underline{Variant 4} 
	&
	\underline{Variant 5}
	& 
	\\
		$
	\begin{array}[t]{l}
	%
	%
	A_{10} \becomes L_{11}^{-1} A_{10} \\
	%
	%
	\rowcolor[gray]{0.8}    
	A_{20} \becomes A_{20} - L_{21} A_{10} ~~~ \mbox{({\sc gemm})} \\
	%
	%
	\rowcolor[gray]{1.0}  
	A_{11} \becomes L_{11}^{-1} A_{11} L_{11}^{-H} \\
	%
	%
	Y_{21} \becomes L_{21} A_{11} \\
	%
	%
	A_{21} \becomes A_{21} L_{11}^{-H} \\
	A_{21} \becomes W_{21} = A_{21} - \frac{1}{2} Y_{21} \\
	%
	%
	\rowcolor[gray]{0.8}    
	A_{22} \becomes A_{22} \\
	\rowcolor[gray]{0.8} 
	~~~ - ( L_{21} A_{21}^H + A_{21} L_{21}^H ) ~~~ \mbox{({\sc her2k})} \\
	\rowcolor[gray]{1.0} 
	A_{21} \becomes A_{21} - \frac{1}{2} Y_{21}  \\
	\end{array}
	$
	&
	$
	\begin{array}[t]{l}
	A_{11} \becomes L_{11}^{-1} A_{11} L_{11}^{-H} \\
	Y_{21} \becomes L_{21} A_{11} \\
	A_{21} \becomes A_{21} L_{11}^{-H} \\
	A_{21} \becomes W_{21} = A_{21} - \frac{1}{2} Y_{21} \\
	\rowcolor[gray]{0.8}    
	A_{22} \becomes A_{22} 
	\\
	\rowcolor[gray]{0.8} 
	- ( L_{21} A_{21}^H + A_{21} L_{21}^H )
	 ~~~ \mbox{({\sc her2k})} \\
	\rowcolor[gray]{1.0}
	A_{21} \becomes A_{21} - \frac{1}{2} Y_{21}  \\
	\rowcolor[gray]{0.8}    
	A_{21} \becomes L_{22}^{-1} A_{21} ~~~ \mbox{({\sc trsm})}
	\end{array}
	$
	&
	
\end{tabular}
	}

\begin{figure*}[tbp]

\begin{center}
    \FlaAlgorithmNarrow
\end{center}

\caption{All five blocked variants for computing $ A \becomes L^{-1} A L^{-H} $.
}
\label{fig:algs}
\end{figure*}

%% file: discussion.tex
{\bf I don't like this section.  Skip to the next section.  This one probably needs to go!}

Formal methods for the verification and derivation of algorithms was seen as critical to the success of computer science as a science in the 1960s and 1970s.  It featured prominently in discussions at the NATO Software Engineering Conference 1969~\cite{Buxton:1970:SET:1102021} as well as Dijkstra's Turing Award lecture, deriving a proof of correctness hand-in-hand with the derivation of algorithms is key to adding value for the software developer.   Yet, we believe experience with these techniques in the field of dense linear algebra software to be the first truly successful application, at least when it comes to deriving loop-based algorithms.  Let us discuss what it is that has made us successful.

\subsection{A community embraces abstraction}

A key development in the right direction, in the 1990s, was the introduction of the Message-Passing Interface (MPI)~\cite{MPI1,MPI2}.  Driven by a need to manage complexity in the new era of distributed memory computing, MPI showed how low-level details about the communication on such architectures could be hidden using object-based programming and opaque objects.  The crisis at the time was sufficiently great that the community embraced this relatively new way of programming.  Rather than directly using  indices when describing the nodes involved in a communication, communicators hid such details.  Similarly, derived data types could be used to hide the details of the storage of data.  By programming closer to the level of abstraction at which one reasoned, the quality of code improved even though formal methods were not applied.  The lesson: embrace abstraction!

\subsection{A disconnect between notation for theory and notation for expressing algorithms/code}

In the field of dense linear algebra, theoretical results were often presented by partitioning matrices into submatrices and algorithms were often explained by arguing about the contents of submatrices~\cite{GVL,GVL4}.  {\bf Check if Wilkinson already did this.}  Yet (in the same works) when algorithms were presented, explicit indexing into arrays was utilized and/or algorithms were explained with index ranges (using MATLAB-like notation).  In the late 1990s, we started to, on the one hand, present algorithms with a notation similar to what we use in this paper (the first journal paper being~\cite{}) and, on the other hand, represent such algorithms in code for distributed memory architectures using object-based techniques that resulted in the code closely resembling the algorithm, as part of the PLAPACK library~\cite{}.  The complexity of describing distribution of computation and data necessitated this.  The lesson: your notation and Application Programming Interface (API) should closely resemble how you reason, hiding as much as possible distracting details.

%% file: questions.tex
\subsection{How broadly has it been applied?}

The methodology has been applied to a broad set of operations that are part of the BLAS and LAPACK, as well as BLAS-like and LAPACK-like operation.  All in all, given how many special cases there are for each operation (e.g., whether the upper or lower triangular part of a matrix is to be updated in the two-sided TRSM), this means it has been applied to hundreds of operations, yielding families of algorithms for each.

Over the years, a number of question have been raised about the scope:  Does it apply to common but more difficult operations like LU with partial pivoting and Householder transformation based QR factorization?  The answer is ``yes''~\cite{TSoPMC}.  
Does it apply to eigenvalue problems?  The answer is ``no yet'' since it has been proven that, in general, there is no algorithm for solving such problems in a finite number of computations. As the example in Section~\ref{sec:derivation} shows, important steps encountered along the way do lend themselves to the technique.  
It may be that the iterative methods that underlie eigenvalue problem solvers can also be developed, but this is an open question.

\subsection{What is the practical impact?}

These algorithms have been incorporated, in one form or another, in our BLAS-like Library Instantiation Software (BLIS)~\cite{BLIS1,BLIS2} and {\tt libflame} libraries~\cite{libflame_ref,Zee:2009:LLD:1674531.1674626}.  Many have also made their way into the Elemental library for distributed memory dense matrix operations.

\subsection{Can formal derivation be easily mastered?}

Proving simple programs correct and deriving such simple programs was long a core part of computer science curricula (but less so more recently)~\cite{Gries}.
For almost two decades, we have taught the practical techniques at the core of this paper in an upper division undergraduate course titled ``Programming for Correctness and Performance.'' We have frequently drawn undergraduates (often as early as in their freshman year) into research on scientific computing by introducing them to these techniques.  More recently, we have taught these techniques as part of a Massive Open Online Course (MOOC) titled ``LAFF-On Programming for Correctness''~\cite{LAFF-On,LAFF-On-edX}.
In our experience, it is easier to get novices  than experienced scientific software developers to embrace formal derivation.  As such, the described methodology democratizes the development of linear algebra algorithms and libraries. 

\subsection{If it is systematic, can it be made automatic?}

Computer science is in part about taking knowledge and making it systematic to the point where a computer can perform related tasks.  Given that the methodology we described in Section~\ref{sec:derivation} is systematic, can it be automated?  The answer is that it can, and that it has been.  As soon as we formulated the described eight steps as the ``worksheet'' given in Figure~\ref{fig:WS}, we recognized it could be automated.  A first prototype was given in the dissertation of Bientinesi~\cite{Paolo:PhD} and the latest efforts in the dissertation of Fabregat~\cite{Fabregat-Traver2014:278}, including an open source tool, Cl1ck%
\footnote{
Source code for Cl1ck available from \href{https://github.com/dfabregat/Cl1ck}{\url{https://github.com/dfabregat/Cl1ck}}.}.  A number of papers  also describe their efforts~\cite{Fabregat-Traver2011:54,Fabregat-Traver2011:238}.

\subsection{How do we choose the best?}

For the example used in Section~\ref{sec:derivation}, we discussed how a distributed memory implementation should ideally cast most computation in terms of ( general or Hermitian) rank-k updates.  The reason is that these operations parallelize well.  In other situations, casting most computation in terms of other operations (e.g., matrix-panel multiplication, for which the bulk of data is read but not written) may be more appropriate.  
One strategy is to derive and implement all algorithms and to then choose the one that attains the best performance.
A good case study for this focuses on inversion of a symmetric positive-definite (SPD) matrix~\cite{Bientinesi:2008:FAR:1377603.1377606}.  Another strategy is to leverage expert knowledge.

In yet another dissertation, Low~\cite{phd:low} shows that a property of the resulting algorithm can be an input to the derivation process.  In this case, he shows, the loop invariant for the algorithm with the desired property can be recognized from the PME, making it only necessary to derive the corresponding algorithm with the described techniques.  In the dissertation, it is argued that this side-steps the so-called phase ordering problem for compilers.

\input dxt

\subsection{What about roundoff error?}
Correctness for scientific computing has a different meaning due to the presence of roundoff error when floating point arithmetic is employed.
If the algorithm is not correct in exact arithmetic, it likely has problems when floating point arithmetic is used.  Thus, the methodology yields a family of algorithms that are candidates.

An algorithm is numerically stable if the computed solution is the exact solution to a slightly modified (perturbed) problem.
Since there tends to already be error in  data and/or entering data as floating point numbers itself introduced error, the best one can hope for is the solution to a slightly changed input.
The goal for numerical algorithms thus is deriving numerically stable algorithms.

A backward error analysis is often the vehicle by which it is shown  that an algorithm is numerically stable.
In the dissertation of Paolo Bientinesi~\cite{Paolo:PhD} and a related paper~\cite{Bientinesi:2011:GMS:2078718.2078728},
it has been shown that the methodology for deriving algorithms in Section~\ref{sec:derivation} can be extended to systematically derive the backward error analysis of matrix algorithms.
The process of deriving numerically stable algorithms is pursued in two steps: first identify algorithms that are correct in exact arithmetic and then derive their backward error analyses.
The methodology is simple enough that it has been successfully taught to undergraduates and beginning graduate students using a technical report version~\cite{FLAWN33} of~\cite{Bientinesi:2011:GMS:2078718.2078728} that includes exercises. 

\subsection{Does it apply beyond dense linear algebra?}
The methodology has been successfully applied to the Krylov subspace methods that underlie (sparse) iterative methods for solving linear systems~\cite{Eijkhout20101805}.  The key there is that by thinking of the vectors that are computed as columns in a matrix, the problem can again be formulated to involve dense matrices.  Very recently, the method has been applied to the computation of all triangles in a graph~\cite{Low:Correctness2017}.  Again, the key is to reformulate this as a matrix computation.
In~\cite{phd:low}, it is argued that the methodology should be applicable to the class of primitive recursive functions.

The general insights are as follows:
The likely target domain should inherently benefit from casting computation as a loop-based algorithm rather than a recursive computation.  In this case:
Start with a notation that captures how one reasons at a high level
(in our case, it is the partitioning of the matrices that is captured by the notation);
Cast the operation to be computed in terms of a recursive definition over an inductively defined data structure.
(In our case, this is the PME.);
Extract from this recursive definition one or more loop invariants.
If the methodology applies, the loop should now naturally follow.

%% file: dxt.tex
Our own approach to implementation of high-performance linear algebra operations breaks the process up into phases.  The first phase is to derive a family of algorithms.  These tend to be loop-based so that the algorithmic block size naturally matches the cache sizes in a hierarchical memory.  The second phase is to optimize the loop body.  For this, we often merge data movements required to implement the various updates in the loop body.  

The operations in a loop body can be viewed as a directed acyclic graph (DAG) where nodes represent computation and/or data movement and edges represent data dependencies.  Optimizations that an expert applies can then be viewed as taking subgraphs and replacing them with alternative, equivalent subgraphs, in an effort to reduce overhead. 
Each step preserves correctness.  The process defines a design space that can be explored for the best implementation.

As part of a dissertation~\cite{phd:marker} and related  papers~\cite{DxTJournal1,iWAPT13}, this has been formalized as a methodology called Design by Transformation (DxT). The approach is systematic enough that it has been automated in a prototype system, DxTer. 
Details go beyond the scope of this paper.  What is important is that in a study that explores the design space that is created~\cite{PerformanceStairs}, it is shown that the design decision that has the greatest impact on performance is what algorithm from the family of algorithms to use.

%% file: conclusion.tex
Had someone told us two decades ago that it was possible to apply goal-oriented programming techniques to dense linear algebra operations, we would have been skeptical. 
Only further research can help explore how far the approach can be pushed.
Thus, one novel contribution of this paper is that it gives evidence that formal derivation of programs is important to software correctness for HPC applications~\cite{DBLP:journals/corr/GopalakrishnanH17}.